 \documentstyle[prl,aps]{revtex}  
\def\bra{\langle} \def\ket{\rangle} \def\ack{\,|\,}

\begin{document}
\draft


 \twocolumn[\hsize\textwidth\columnwidth\hsize  
 \csname @twocolumnfalse\endcsname              

\title{
$g$-Factors and the  
Interplay of Collective and Single-Particle Degrees of Freedom \\
in Superdeformed Mass-190 Nuclei
} 
\author{Yang Sun$^{1,2,3}$, Jing-ye Zhang$^1$,
Mike Guidry$^1$
}
\address{
$^1$Department of Physics and Astronomy, University of Tennessee,
Knoxville, Tennessee 37996 \\
$^2$Department of Physics, Tsinghua University, Beijing 100084, P.R. China\\
$^3$Department of Physics, Xuzhou Normal University,
Xuzhou, Jiangsu 221009, P.R. China\\
}

\date{\today}
\maketitle

\begin{abstract}
Interplay of collective and single-particle degrees of freedom
is a common phenomenon in strongly correlated many-body systems.  
Despite many successful efforts in the study of superdeformed
nuclei, there is still unexplored physics that can be best understood only
through the nuclear magnetic properties. 
We point out that study of the gyromagnetic factor (g-factor) may open a unique
opportunity for understanding superdeformed structure.  
Our calculations suggest that investigation of the g-factor dependence 
on {\it spin and particle number} can provide 
important information 
on single-particle structure and its interplay with collective motion 
in the superdeformed mass-190 nuclei.
Modern experimental techniques 
combined with the new generation of sensitive detectors
should be capable of testing our predictions.
\end{abstract}

\pacs{21.10.Ky, 21.10.Re, 21.60.Cs, 27.80.+w}

 ]  

\narrowtext

The study of superdeformed nuclei has been an active focus of 
low-energy nuclear physics. 
The richness of the 
experimental data \cite{Han99} has motivated many theoretical
efforts. Most of the superdeformed yrast spectra can be
well described by several models.  
For the mass-190 region, it is found that, even though these nuclei are
among the most deformed, they exhibit substantial deviation
from rigid rotor behavior as the angular momentum increases, 
and the deviation varies from nucleus to nucleus. 
This deviation is a microscopic consequence of interplay between
collective and single-particle motion in a nuclear many-body system. 
The second well-studied quantity besides the $\gamma$-ray energies, the
transitional quadrupole moment, is insensitive to particular
contributions from individual neutron and proton orbitals.   

In a recent Letter \cite{Sun97}, 
we have presented a systematic description of yrast superdeformed bands in
the mass-190 region. The theoretical analysis was based on the projected
shell model (PSM) \cite{HS95}. 
Without any special adjustment in parameters,  
excellent agreement with available spectroscopic and transition 
quadrupole moment 
data for the series of even-even Hg and Pb isotopes was obtained.  
Quantitative predictions were also made for g-factors in $^{194}$Hg in
which explicit dependence of the g-factors on spin was given. 

The g-factor is 
sensitive to the single-particle structure in wave functions
as well as to its interplay with collective degrees of freedom.
Because of the intrinsically opposite signs of the neutron and proton
$g_s$,
a study of g-factors enables determination of the microscopic structure
for underlying states. For example, variation of g-factors
often is a clear indicator for a
single-particle component that strongly influences the total wave function.
One can extract g-factors from experimental 
$B(M1)/B(E2)$ ratios, but this usually involves 
introduction of several assumptions. 
Therefore, direct measurement of g-factors is most valuable, 
though this is a very challenging task. 
  
Considerable progress has been made recently in g-factor measurement 
\cite{g-factor}. 
By using the transient field method combined with 
the high-resolution detector Gammasphere, the first direct 
measurement
of g-factors in the superdeformed well for $^{194}$Hg has been
reported in a recent Rapid Communication \cite{g-hg194}.
In this paper, Mayer {\it et al.} concluded 
that the measured g-factors are in agreement
with the theoretical predictions of 
Hartree-Fock plus BCS calculations \cite{Per97},
as well as with the picture of a rigid 
rotation for which $g=Z/A$ (where $Z$ and
$A$ are proton and total particle number, respectively). 

In the present paper, we shall point out that (1) 
g-factors of superdeformed bands 
in even-even nuclei of the mass-190 region  
generally behave very differently from the collective $Z/A$ estimates, 
and variation of g-factors {\it as a function of spin} can provide us
with information on the evolution of pair correlations; 
(2) {\it As a function of particle number}, 
our calculation shows another variation of g-factors
reflecting shell fillings, 
which indicates important
contribution from single-particles at different rotational stages.
Both variations 
contain valuable information on superdeformed structure and 
certainly suggest a picture beyond rigid body rotation. 
However, current experimental uncertainties \cite{g-hg194} 
are still too large to 
distinguish such features and thus do little to test theories. 

Our calculation is performed by using the PSM, in which the same 
separable-force Hamiltonian and same configuration space are employed as 
in our previous Letter \cite{Sun97}. 
To treat these heavy, superdeformed nuclei, we include in our single-particle
configuration space 4 major shells for each kind of nucleon, 
i.e. $N=4, 5, 6, 7$ for neutrons and
$N=3, 4, 5, 6$ for protons. 
In the PSM, 
the many-body wave function is a superposition of
(angular momentum) projected multi-quasiparticle states,
\begin{equation}
| \psi^I_M \rangle ~=~
\sum_{\kappa} f_{\kappa} \hat P^I_{MK_\kappa}
| \varphi_{\kappa} \rangle ,
\label{ansatz}
\end{equation}
where $\hat P^I_{MK}$ is the projection operator \cite{RS80} and 
$| \varphi_{\kappa} \rangle$ denotes basis states consisting of
the quasiparticle (qp) vacuum, 2 quasi-neutron and -proton, and
4-qp states 
\begin{eqnarray}
\ack\varphi_\kappa\rangle=\left\{ \ack 0 \rangle,
\ a^\dagger_{\nu_1}a^\dagger_{\nu_2}\ack 0\rangle,
\ a^\dagger_{\pi_1}a^\dagger_{\pi_2}\ack 0\rangle,
\ a^\dagger_{\nu_1}a^\dagger_{\nu_2}a^\dagger_{\pi_1}
a^\dagger_{\pi_2}\ack 0\rangle \right\}.
\label{conf}
\end{eqnarray}
In Eq. (\ref{conf}), 
$a^\dagger$'s are the qp creation operators, $\nu$'s ($\pi$'s)
denote the neutron (proton) Nilsson quantum numbers which run over
properly selected (low-lying) orbitals and $\ack 0 \rangle$ the
qp vacuum or 0-qp state.
The dimension of the qp basis
(\ref{conf}) is about 100 and the deformation of the basis is 
fixed at  
$\epsilon_2 = 0.45$ for all nuclei calculated in this paper.  
The choice of this deformation is based on experimental information
\cite{Han99}. 
We note that the choice of a basis deformation is not very critical  
for the PSM because many dynamical fluctuations 
are taken into account by  
the configuration mixing. 
The same Hamiltonian (including spherical single-particle, 
residual quadrupole--quadrupole, monopole pairing, and quadrupole
pairing terms) as employed in Ref. \cite{Sun97} is then 
diagonalized in (\ref{ansatz}).  
The resulting energy spectrum has been compared with data 
as presented in \cite{Sun97},
and the wave function $| \psi^I_M \rangle$ is used to
calculate g-factors:  
\begin{equation}
 g(I) = {{\bra\psi^I_{M=I} | \hat\mu_z | \psi^I_{M=I}\ket} \over {\mu_N I}}  
      = {{\bra\psi^I || \hat\mu || \psi^I\ket} \over {\mu_N \sqrt{I(I+1)}}}, 
\end{equation}
with $\hat\mu$ being the magnetic vector 
and $\mu_N$ the nuclear magneton. 
$g(I)$ can be written as a sum of the proton and neutron parts 
$g_\pi(I) + g_\nu(I)$, with 
\begin{eqnarray}
g_\tau(I)& = & \frac{1}{\mu_N \sqrt{I(I+1)}} 
\nonumber \\ && \times \left(
     g^{\tau}_l \bra \psi^I||\hat j^\tau ||\psi^I \ket 
     + (g^{\tau}_s - g^{\tau}_l) \bra\psi^I||\hat s^\tau||\psi^I\ket \right),
\label{g-form}
\end{eqnarray}
where $\tau= \pi$ or $\nu$, 
and $g_l$ and $g_s$ are the orbital and spin
gyromagnetic ratios, respectively.
We use the free values for $g_l$ 
and the free values damped by the usual 0.75 factor for $g_s$ 
\cite{BM75} 
\begin{eqnarray}
g_l^\pi &=& 1  \;\;\;\;\;\;\; g_s^\pi = 5.586 \times 0.75 
\nonumber \\
g_l^\nu &=& 0  \;\;\;\;\;\;\; g_s^\nu = -3.826 \times 0.75 .
\end{eqnarray}
We emphasize that
g-factor in the PSM is computed directly from 
the many-body wave function (Eqs. (\ref{ansatz}) -- (\ref{g-form})) 
without a semiclassical separation of the collective 
and the single-particle parts. 
In particular, 
there is no need to introduce a core
contribution, $g_R$, which is a model-dependent concept
and not a measurable quantity.
Previous g-factor calculations of the PSM have been tested by
experiment. For example, predictions for the normally deformed 
rare earth nuclei \cite{HS95,SE94} 
were supported by later measurement \cite{g-Er}. 

Our calculated $g$-factors for $^{194}$Hg (filled diamonds) 
are shown in Fig.\ 1.  
In the low spin range, 
we observe a smooth increase in the $g$-factors up to the spin states
around $I = 26$. 
Bengtsson and \AA berg suggested that an increase in g-factor
at low spins was due to changes in deformation and pairing \cite{Ben86}.
Since deformation does not have a large variation along a band 
in this heavy, well-superdeformed nucleus 
($<2\%$ as seen from our calculation in 
\cite{Sun97} and from relativistic 
\cite{AKR99} and non-relativistic \cite{VER00} mean field theories), 
the g-factor increase in the low spin range can be attributed to 
the pairing change caused by the Coriolis antipairing
effect, which coherently weakens pair correlation across many pairs.  
The weakened pairing was demonstrated in our calculations for pairing gaps 
in \cite{Sun97} and was suggested as the main source 
for the increase in the moment of inertia
before $I = 26$. 

Above the spin $I = 26$, an accelerated increase is seen for 
the range $I = 26 - 44$. At $I =
44$, the $g$-factor reaches a value of $0.41 \approx Z/A$. 
This behavior suggests that rotation
alignment of high-$j$ orbitals plays an additional role in this spin range,
adding contributions to the g-factor increase in addition to 
that from the coherently
weakened pairing. 
This is consistent with the mechanism that 
we suggested \cite{Sun97} to explain the 
observed gradual increase in dynamical moment of inertia 
in superdeformed mass-190 nuclei. 
The rotation
alignment of high-$j$ orbitals enhances the moment of inertia, but
the high-$j$ alignment contribution alone seems insufficient
to cause the pronounced increase in the moment of inertia seen in 
$^{194}$Hg. In particular, this cannot easily
explain the increase in the moment of inertia 
before $I = 26$, where the high-$j$ pair alignment is negligibly small.
This combined effect of high-$j$ alignment and decrease of pairing
is also seen in cranked relativistic
Hartree-Bogoliubov theory \cite{AKR99}. 

The g-factor measured by Mayer {\it et al.} \cite{g-hg194} is plotted in
Fig. 1 for comparison. 
Because the error bar is large, it is impossible to draw 
a definite conclusion. 
However, it is interesting to note that 
the experimentally suggested spin range for this measurement  
is such that the rotation
alignment of high-$j$ orbitals is just beginning to contribute. 
Therefore, the current measurement lies in a sensitive spin range 
where one could have a chance to track the competition of 
the Coriolis antipairing and the high-$j$ pair alignment. 
A test of these ideas will require reduction of the experimental 
uncertainty by a factor of at least $2-3$,
and extension of the data to at least spin states $I=36-42$. 

Collective g-factors were calculated systematically by Perries {\it et al.}
\cite{Per97} 
using the Hartree-Fock plus BCS theory with the $SkM^*$ effective force.
Since the dependence of g-factors on spin was not given in their
calculation, the results may be considered as a band average over 
the low spin states.  
The g-factor obtained for $^{194}$Hg by Perries {\it et al.} \cite{Per97}
is also shown in Fig. 1 for comparison (dashed line). 
It is obviously very different from the $Z/A$ value (dashed-dotted line), 
which was the main conclusion and was clearly stated in \cite{Per97}.  
The value of Perries {\it et al.} matches well with our calculations 
for the spin range from the measured bandhead up to the $I=26$ state. 
However, the two calculations diverge after that spin as the PSM results 
receive contributions from the high-$j$ pair alignment. 

To see this clearly, we have made a test calculation with 
the PSM in which all the multi-qp
configurations in Eq. (\ref{conf}) are excluded except the vacuum (0-qp state). 
This effectively removes contributions of any pair alignment to
the total wave function. The results are shown in Fig. 1 as open squares.  
As can be seen, they are close to that of Perries {\it et al.}, although
a visible increase along the band still exists in the 
PSM calculation which is the result caused by
the Coriolis antipairing effect as discussed before.  
However, these results are not from our complete shell model basis
and the important ingredients in the basis, which can correctly describe
the higher spin behavior of the moments of inertia including the 
observed downturning \cite{Cedar94} at the highest spins,
are now excluded. 
It is seen, from the PSM results in Fig. 1, 
the rigid body g-factor value $Z/A$ in this case can only be reached 
at the highest spin states when 
the full space in Eq. (\ref{conf})  
is employed. 

Thus, although these heavy nuclei have large deformations and
the deformation is very stable against rotation, they are not  
simple rigid rotors.
Because this can not be seen obviously from the 
spectroscopy and
transition quadrupole moments,
accurate measurement of g-factors may play a unique role in exploring 
the microscopic structure, as we shall further demonstrate below. 

It is well-known that 
high-$j$ particles which lie in orbits close to Fermi levels 
and most easily align their spins along the axis of system rotation 
may have   
significant influence on g-factors if the pair correlation 
among them is broken. 
In the superdeformed well of the mass-190 Hg and Pb nuclei, 
$K = {3\over 2}$ and $5\over 2$ protons in the $i_{13/2}$ orbital,
and $1\over 2$, $3\over 2$ and $5\over 2$ neutrons in the 
$j_{15/2}$ orbital 
are such particles. 
We expect that changing the single-particle environment around the Fermi
levels for different shell filings will lead to different pattern of 
g-factors.  
Fig. 2 shows the g-factors of superdeformed yrast bands
calculated by the PSM for the Hg and Pb isotopes 
with neutron numbers from 108 to 116. 
The wave functions used to calculate these g-factors are obtained 
when solving the eigenvalue equations, with the corresponding energies
having been presented in Figs. 1 and 2 of Ref. \cite{Sun97}.  

Three observations can be made immediately from Fig. 2:
(1) For all nuclei presented, the g-factors clearly deviate
from the corresponding rigid body value $Z/A$. For the Hg isotopes, 
there are narrow spin ranges where the g-factors can be close to $Z/A$
(but such spin ranges are very distinct in different isotopes). 
However, for the Pb isotopes, the g-factors are systematically
smaller than the corresponding rigid body value for all 
spin states calculated.  
We note that a smaller g-factor for Pb isotopes relative to Hg  
has also been seen in the effectively spin-averaged calculations of
Perries {\it et al.} \cite{Per97}. 
(2) For the spin range where g-factors in $^{194}$Hg were measured,
a clear increase in g-factors as a function of spin is predicted for
the Hg isotopes, but in Pb, they are nearly constant.   
The rate of increase in the Hg isotopes is obviously different,
with the smallest seen in $^{192}$Hg. 
(3) A drop in g-factors is predicted at high spins ($I=44 - 48$) 
for the lightest isotopes. 

These results can be understood  
as consequences of competition between the high-$j$ 
neutron and proton particles. 
Note that the yrast band is obtained by band mixing from the
unperturbed bands with the lowest one in energy having the 
largest contribution. 
By analyzing the wave functions, we find that 
the 2-quasiproton band ($\pi i_{13/2} [{3\over 2}, {5\over 2}]$, $K = 1$) 
is lower in the Hg isotopes, and thus closer to the 0-qp band.
This increase in the importance of proton components 
makes the g-factors in the Hg's systematically
larger than those in the Pb's. As spin increases, this
2-quasiproton band 
approaches the 0-qp band, leading the g-factor increase.  
However, depending on the positions of 2-quasineutron bands, 
influence of the 2-quasiproton band can be largely compensated if sufficient 
single-neutron components are mixed in the wave function. 
The relevant 2-quasineutron bands have the structure 
$\nu j_{15/2} [{1\over 2},
{3\over 2}]$ and $\nu j_{15/2} [{3\over 2},
{5\over 2}]$, both coupled to $K = 1$. 
For lighter isotopes with neutron numbers $N=108$ and 110, these two  
2-quasineutron bands approach the 0-qp band at higher spins and 
compete with the 2-quasiproton band,  
thus can have large influence in the wave function for the higher spin states. 
This leads to a drop in g-factors at $I= 44-48$. 
For the heavier isotopes, however, only one 2-quasineutron band
($\nu j_{15/2} [{3\over 2}, {5\over 2}]$) is low enough in energy 
to compete with the 2-quasiproton band.
Therefore, the g-factor drop at high spins in the heavier isotopes 
is much diminished. 

In Table I, 
we take $^{190}$Hg as an example and 
show the coefficients $f_\kappa$ in the yrast wave function 
(see Eq. (\ref{ansatz}))  
for four low-lying configurations that strongly 
govern the evolution of g-factors.  
While the 0-qp band dominates absolutely the wave function for low
spin states,  
the 2-qp components become more important  
in the wave functions as spin increases.  
Comparing the relative size of the numbers at each spin, 
we see a much faster increase of the 2-quasiproton band. This 
suggests a dominant role for the protons over the neutrons in the states
with low and intermediate spins. 
For higher spin states, the combined contribution of the two
2-quasineutron bands becomes important and eventually this dominates the
wave function for states at the highest spins. 

We conclude that  
experimental information on the g-factor would be very  
valuable for exploring the underlying microscopic inside 
of the superdeformed nuclei. 
Unfortunately, 
the currently measured g-factor \cite{g-hg194} is not accurate enough 
to make a meaningful comparison with theory. 
As we can see in Fig. 2, 
the experimental 
error bar extends over the theoretical predictions 
in all the five Hg isotopes, 
and thus can make no distinction for the underlying
microscopic structure discussed above.    
To begin to make meaningful comparisons, the g-factor measurement 
uncertainties must be reduced by at least a factor of $2-3$, and 
measurements extended to higher spin states. 

In summary, 
our understanding of the microscopic structure for the superdeformed 
mass-190 nuclei is not complete and the quantities that can
distinguish physics as functions of spin and particle number have not been 
well studied.  
The g-factor is such a quantity which contains information on the 
interplay of collective and single-particle degrees of freedom. 
The problem discussed here specifically for mass-190 nuclei may
be common for other superdeformed
mass regions. 
New experimental techniques such as those discussed in Ref. \cite{g-factor}
hold the promise of permitting these prospects to be tested thoroughly 
for the first time. 

Communication with Prof. N. Benczer-Koller is acknowledged. 
Y.S. thanks Prof. Gui-Lu Long of Tsinghua University
for warm hospitality, and for support from
its senior visiting scholar program.
                            Research at the University of
                            Tennessee is supported by the U.~S. Department
                            of Energy through Contract No.\
                            DE--FG05--96ER40983.

\baselineskip = 14pt
\bibliographystyle{unsrt}

\begin{figure}
\caption{
g-factors for the superdeformed yrast band of $^{194}$Hg.
Filled diamonds are the PSM results with the full model space, open squares  
the PSM results calculated in the projected 0-qp state only, 
filled circle (with error bars) the experimental data \protect\cite{g-hg194},
dashed line the HF+BCS result \protect\cite{Per97}, and dashed-dotted line
the rigid body value. 
}
\label{figure.1}
\end{figure}

\begin{figure}
\caption{
Predicted g-factors for superdeformed yrast band of $^{188-196}$Hg and
$^{190-198}$Pb. 
}
\label{figure.2}
\end{figure}

\begin{table}[h]
\begin{center}
\caption{
Wave functions $f_\kappa$ (see Eq. (\protect\ref{ansatz})) 
for the superdeformed yrast band of $^{190}$Hg. 
Only components of the 0-qp, 2-qp proton band  
($\pi i_{13/2} [{3\over 2}, {5\over 2}]$, $K = 1$),
2-qp neutron band I ($\nu j_{15/2} [{1\over 2}, {3\over 2}]$, $K = 1$), 
and 2-qp neutron band II ($\nu j_{15/2} [{3\over 2}, {5\over 2}]$, $K = 1$)
are shown. 
Note that, because of non-orthogonality of the projected basis, 
only the relative size of the numbers for a given spin is significant. 
}
\vspace{0.5cm}
\begin{tabular}{c|c|c|c|c}
Spin $I$ & 0-qp band & 2-qp $\pi$ band & 
2-qp $\nu$ band I & 2-qp $\nu$ band II \\
\hline
10&0.4235&0.0285&0.0315&0.0119\\
14&0.3692&0.0366&0.0349&0.0155\\
18&0.3666&0.0505&0.0396&0.0217\\
22&0.3980&0.0743&0.0454&0.0326\\
26&0.4589&0.1154&0.0521&0.0522\\
30&0.5452&0.1857&0.0595&0.0875\\
34&0.6472&0.3017&0.0697&0.1503\\
38&0.7499&0.4827&0.0936&0.2613\\
42&0.8343&0.7510&0.1649&0.4600\\
46&0.8450&1.1046&0.3718&0.8117\\
50&0.6607&1.4235&0.8324&1.3317\\
54&0.2782&1.7152&1.5114&2.0016\\
\end{tabular}
\end{center}
\end{table}

\end{document}